\newif\ifanonymousversion
\anonymousversionfalse

\documentclass[10pt,conference]{IEEEtran}
\usepackage{cite}
\usepackage{amsmath,amssymb,amsfonts}
\usepackage{algorithmic}
\usepackage{graphicx}
\usepackage{textcomp}
\usepackage{xcolor}
\usepackage[hyphens]{url}
\usepackage{soul}
\usepackage{multicol, multirow}
\usepackage{graphicx}
\usepackage{subcaption}
\usepackage[margin=1in]{geometry}
\usepackage{booktabs} %
\usepackage{flushend}

\def\BibTeX{{\rm B\kern-.05em{\sc i\kern-.025em b}\kern-.08em
    T\kern-.1667em\lower.7ex\hbox{E}\kern-.125emX}}

\newcommand{\hpcasubmissionnumber}{2435}

\begin{document}

\title{Trace-Based Reconstruction of Quantum Circuit Dataflow in Surface Codes} 

\ifanonymousversion

\author{{\normalsize{HPCA 2026 Submission
      \textbf{\#\hpcasubmissionnumber} -- Confidential Draft -- Do NOT Distribute!!}}}

\else

\author{
Theodoros Trochatos$^{1}$,
Christopher Kang$^{2}$,
Andrew Wang$^{3}$,
Frederic T. Chong$^{2}$,
Jakub Szefer$^{4}$ \\
\small $^{1}$Yale University, New Haven, CT, USA \\
\small $^{2}$University of Chicago, Chicago, IL, USA \\
\small $^{3}$Cornell University, Ithaca, NY, USA \\
\small $^{4}$Northwestern University, Evanston, IL, USA
}

\fi
\maketitle
\thispagestyle{plain}
\pagestyle{plain}

\begin{abstract}
Practical applications of quantum computing depend on fault-tolerant devices that employ error correction. A promising quantum error-correcting code for large-scale quantum computing is the surface code. For this code, Fault-Tolerant Quantum Computing (FTQC) can be performed via lattice surgery, i.e. merging and splitting of encoded qubit patches on a 2D grid. Lattice surgery operations result in space-time patterns of activity that are defined in this work as {\em access traces}. This work demonstrates that the {\em access traces} reveal when, where, and how logical qubits interact. Leveraging this formulation, this work further introduces TraceQ, a trace-based reconstruction framework that is able to reconstruct the quantum circuit dataflow just by observing the patch activity at each trace entry. The framework is supported by heuristics for handling inherent ambiguity in the traces, and demonstrates its effectiveness on a range of synthetic fault-tolerant quantum benchmarks. The {\em access traces} can have applications in a wide range of scenarios, enabling analysis and profiling of execution of quantum programs and the hardware they run on. As one example use of TraceQ, this work investigates whether such traces can act as a side channel through which an observer can recover the circuit’s structure and identify known subroutines in a larger program or even whole programs. The findings show that indeed the minimal {\em access traces} can be used to recover subroutines or even whole quantum programs with very high accuracy. Only a single trace per program execution is needed and the processing can be done fully offline. Along with the custom heuristics, advanced subgraph matching algorithms used in this work enable a high rate of locating the subroutines while executing in minimal time.

\end{abstract}

\section{Introduction}
\label{introduction}

Fault-tolerant quantum computers can reliably perform quantum computations, despite  the presence of inherent hardware noise andimperfections inherent. Quantum computer hardware is projected to have  physical error rates ranging from $10^{-3}$ to $10^{-4}$~\cite{google2023suppressing, moses2023race}, while quantum programs require logical error rates that are orders of magnitude lower. 
Fortunately, quantum error correction (QEC)~\cite{calderbank1998quantum, gottesman1997stabilizer, dennis2002topological} produces logical qubits with reduced error rates by using additional physical qubits to produce redundancies to faults.
QEC employs multiple physical qubits alongside a classical decoding algorithm to identify and correct physical errors. Large-scale, Fault-Tolerant Quantum Computers (FTQC)  are expected to solve computational problems that are otherwise intractable, e.g., factoring large numbers~\cite{shor1999polynomial} or simulating molecular systems~\cite{zhang2025fault}. 

\begin{figure}[t]
\centering
\includegraphics[width=1\linewidth]{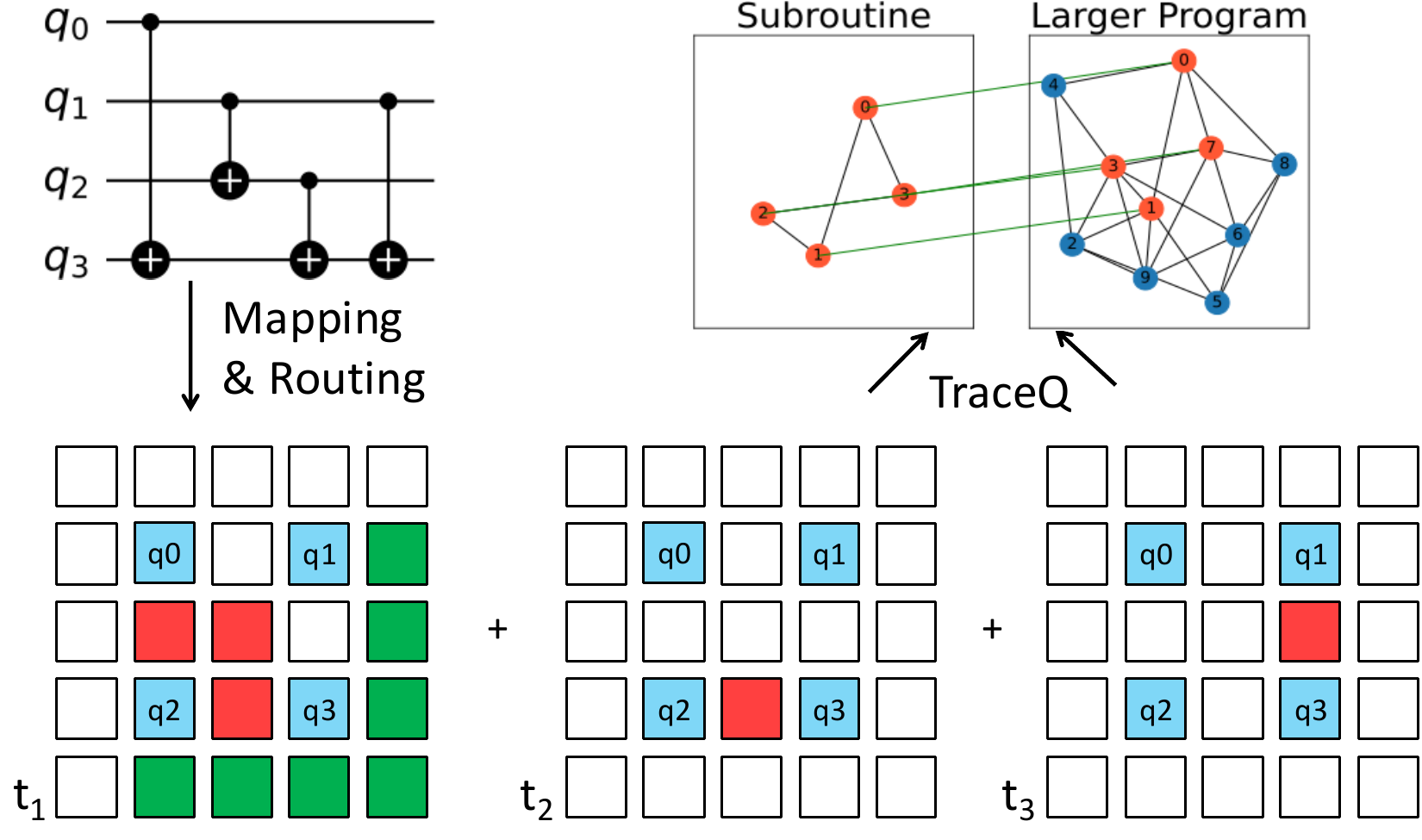}
\caption{\small High-level operation of fault tolerant quantum computer and where TraceQ fits within this operation. Qubits of a quantum circuit (top-left) are mapped and routed (bottom row) onto a 2D grid of logical qubits, also called patches. Routing is performed to realize each quantum gate via lattice surgery which connects patches when gate operation is performed. During circuit execution routing manifests itself as space-time patterns of activity. TraceQ (top-right) recovers quantum circuit dataflow from the activity traces.}
\label{fig_mapping_routing_grid}
\end{figure}

A promising quantum error-correcting code is the surface code~\cite{fowler2012surface, Litinski_2019, litinski2022active}. 
On surface codes, two-qubit gates are effected via \textit{lattice surgery}~\cite{horsman2012surface,fowler2018low}.
In lattice surgery, logical qubits are represented by {\em patches} of the surface code. The most basic type of {\em patch} is a {\em tile}. Tiles can be merged to form larger patches, or reversely, patches can be split to smaller patches~\cite{Tan_2024}.
As shown in Figure~\ref{fig_mapping_routing_grid}, lattice surgery is used to enable fault-tolerant logical operations by the merging and splitting of encoded qubit patches, which are rectangular regions of entangled physical qubits arranged on a 2D lattice. Logical operations such as two-qubit gates or T gates require temporary {\em merging} of these patches, which are later {\em split}. Surface codes are well-suited to existing physical qubit topologies and support universal computation via magic states. 

Lattice surgery manifests itself as space-time patterns of activity in the patches during merging and splitting. As shown in Figure~\ref{fig_mapping_routing_grid}, the activity of patches changes as each set of gates executes. Lattice surgery and resulting activity patterns are significant for various reasons. For example, the activity patterns (which patches are merging or splitting) indicate which (logical) qubits are interacting (for two qubit operations or for when T-states are accessed). Or, the paths that are created during merging and how many are done in parallel reveals both parallelism in the algorithm (how many gates are executing at the same time), but also congestion and other limitations resulting from the underlying hardware (how long the paths have to be, how roundabout their paths are to connect qubits, etc.). {\em We postulate that these space-time patterns of activity resulting from lattice surgery encode all information needed to recover quantum circuit dataflow, circuit operation, and available parallelism on the target quantum computer.}

Consequently, to test this postulate, we propose {\em access traces} as means of capturing the space-time patterns of activity resulting from lattice surgery. The traces are concise representations of the qubit patch activity as a time-series over duration of the circuit. The traces capture binary data about each patch (active or not active) at each time step of quantum circuit's execution. This minimal representation omits all routing paths, and qubit roles or their labels. Such traces can be collected easily by the quantum computer controllers for profiling purposes without requiring any metadata or additional information from the user. Or they could be captured by malicious attackers wanting to learn about quantum circuits or their operation. Given the minimal and concise trace information and representation, this work addresses two key research~questions:

\vspace{0.2em}

\textit{Q1: Can one recover the logical dependencies structure of a quantum program using only binary activity traces from a fault-tolerant execution?}

\vspace{0.2em}

\textit{Q2: Can one efficiently find and identify types of subroutines, or functions, used within the quantum program?}

\vspace{0.2em}

To answer these questions, we introduce TraceQ, a framework designed to reconstruct the quantum circuit dataflow from these minimal activity {\em access traces} from surface codes. 
To describe the dataflow of a quantum circuit, we employ
a Directed Acyclic Graph (DAG), known as a dependency graph.
Our approach attempts to reconstruct logical dataflow DAG of the quantum circuits from space-time activity routing traces. TraceQ incorporates three key ideas: first, we use a set of heuristics to infer qubit locations and gate connectivity just from observing active and idle patches with no initial knowledge of logical qubit location within the patches. Second, we reconstruct a temporally ordered dataflow DAG of two-qubit gates by analyzing path structures and applying per-qubit dependency constraints. Third, we detect known subroutines from a larger program by applying exact subgraph matching to the reconstructed DAGs. 

We evaluate the TraceQ framework using a range of synthetic FTQC benchmarks. From purely binary activity traces, we are able to reconstruct logical two-qubit gate DAGs and identify embedded subroutines even under ambiguity. Our evaluation includes $600$ randomized synthetic circuits and $28$ subroutine quantum circuits, demonstrating the efficiency and resilience of the approach under different layouts and qubit mapping assignments. The main contributions are listed below.

\subsection{Contributions}

\begin{enumerate}
\item We formulate the new problem of quantum program recovery from binary patch-level activity traces in surface-code-based fault-tolerant quantum computing.
\item We present TraceQ, a trace-based reconstruction framework that infers the logical dataflow DAG from ambiguous active or idle patches trace inputs using a set of heuristics.
\item We demonstrate TraceQ’s ability to reconstruct two qubit gate-level structure and identify known subroutines across a wide range of randomized synthetic FTQC workloads.
\item We evaluate the robustness of our approach under three different grid layouts and different mapping strategies.
\end{enumerate}

\section{Background}
\label{background}

This section covers the background of the classical control infrastructure required to operate a quantum computer and the varied environments in which the infrastructure must operate. It also describes the surface code, a widely used code for scalable fault-tolerant quantum computers.

\subsection{Classical Control Infrastructure}

\begin{figure}[t]
  \centering
  \includegraphics[width=0.85\linewidth]{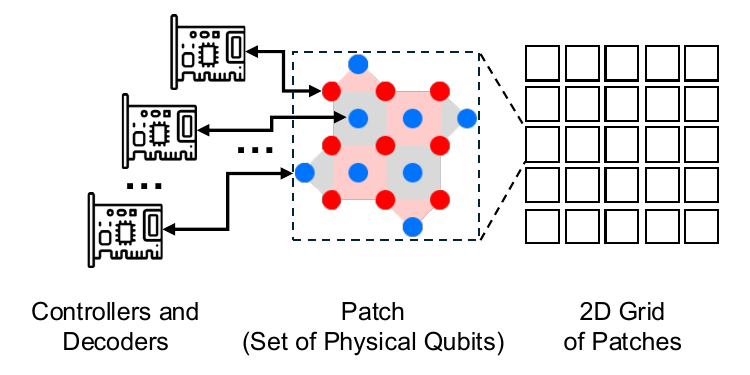}
  \caption{Classical control infrastructure: Controllers, containing QEC decoders and other logic, interface to the physical qubits, each group of physical qubits makes up a patch or a logical qubit.}
  \label{fig_infrastructure}
\end{figure}

Quantum computers depend on classical infrastructure to execute a quantum circuit. Classical control infrastructure must accomplish a range of computationally intensive tasks, from producing electromagnetic pulses which control physical qubits to decoding error syndromes and orchestrating higher level instructions~\cite{zhang2023classical}. 
Some hardware components of the infrastructure include arbitrary waveform generators, FPGAs, and similar equipment which is used to interface with the qubits, as shown in Figure~\ref{fig_infrastructure}. 
These hardware components contain Digital to Analog Converters (DACs) for sending signals to qubits and Analog to Digital Converters (ADCs) for reading out signals from qubits. In addition, hardware such as FPGAs or dedicated ASICs are used to generate the signals going to DACs, and to read out the responses from ADCs. This control infrastructure not only generates control pulses to execute quantum gates on the qubits, but also performs operations such as syndrome decoding and error correction. Such hardware is currently running at room temperature, although there exist proposals to co-locate ASICs within refrigerated quantum chips to perform syndrome decoding~\cite{charbon2016cryo}.

\subsection{Surface Codes, Lattice Surgery, Patch Activity}

\begin{figure}[t]
  \centering
  \includegraphics[width=0.8\linewidth]{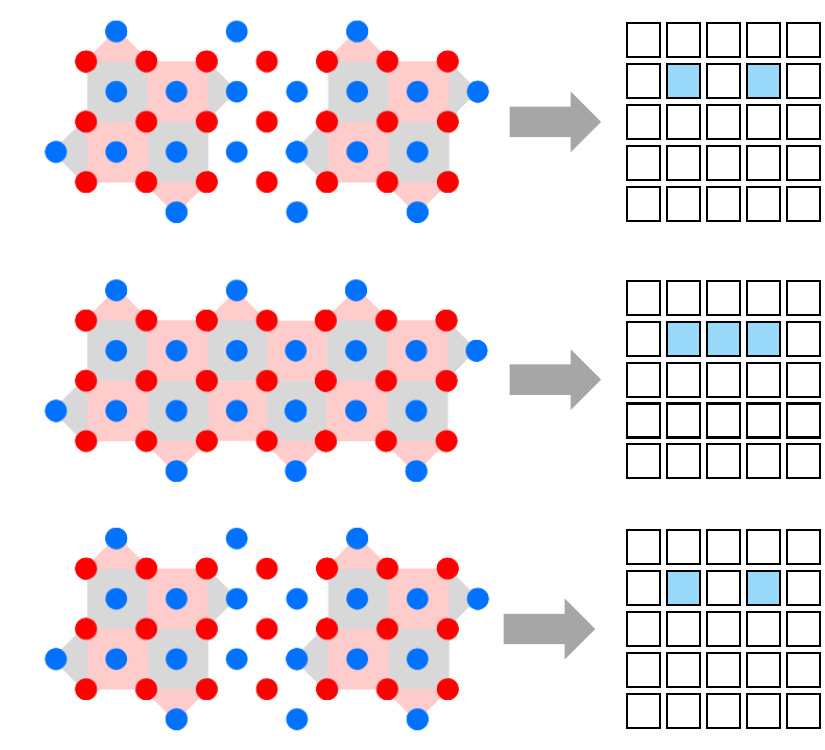}
  \caption{Visualization of a surface code. \textit{Above}: two surface code patches next to each other. The data qubits are red and the syndrome qubits are blue. The stabilizers are shown in shaded regions.
  \textit{Middle}: the patches are merged via lattice surgery. This requires enabling stabilizers in the intermediate region. To decode the merged patch, the entire region must be decoded. \textit{Bottom}: the two patches are split, returning the two patches to the original configuration. \textit{Right}: 2D grid representation of active and idle patches.}
  \label{fig_surface_code}
\end{figure}

QEC is used to produce a logical qubit with lower error rates than its constituent physical qubits. The surface code is a leading QEC code for fault-tolerant quantum computation. As shown in Figure~\ref{fig_surface_code}, a surface code \textit{patch} consists of two types of qubits arranged in a 2D array: \textit{data} qubits that store the logical qubit's state and \textit{syndrome} qubits that are used to detect quantum errors. Syndrome qubits are periodically measured, providing information to a \textit{decoder} which identifies and corrects errors on the data qubits. The decoder is part of the classical hardware depicted earlier in Figure~\ref{fig_infrastructure}.

To compute with surface codes, quantum programs will be expressed in terms of \textit{lattice surgery operations}. These operations involve deforming adjacent surface code patches so that they form one contiguous region. This effects a multi-body Pauli measurement upon multiple logical qubits.  This is different from the typical gate-based model of FTQC, but is equivalent with constant overhead~\cite{fowler2012surface}. Through program execution, different qubits may be accessed. Thus, depending on the program, we would expect different access patterns.

Decoding quantum errors is a classically demanding task and must be executed in \textit{real-time}, i.e., at the cycle rate of physical hardware~\cite{Neven_2024}. Decoding is known to be hard theoretically, with complexity class either NP complete or \#P depending on the precise decoding problem~\cite{iolius2023decoding}. While modern decoding algorithms have improved substantially, they are still expected to require dedicated classical hardware~\cite{barber2025real} shown earlier in Figure~\ref{fig_infrastructure}. Furthermore, during lattice surgery, decoding can increase in complexity as multiple regions must be decoded in tandem~\cite{chamberland2022universal}.

For our work, we assume the most basic case where decoding is performed via room-temperature ASICs. We also assume each surface code patch has a dedicated chip to decode it. This is especially relevant for large surface codes: at distances exceeding $d > 11$, there are already hundreds of syndrome qubits to measure and decode, a significant challenge for decoding chips.

Figure~\ref{fig_surface_code} (right side) also shows the patch-level activity as the surface code executes. Merging and splitting of patches is observed as changes in active and idle patches. Our {\em access traces} capture this activity, which is analyzed by our TraceQ framework.

\section{Access Traces}
\label{problem_formulation}

\begin{figure*}[t]
  \centering
  \includegraphics[width=0.9\linewidth,trim={0cm 0cm 0cm 0cm},clip]{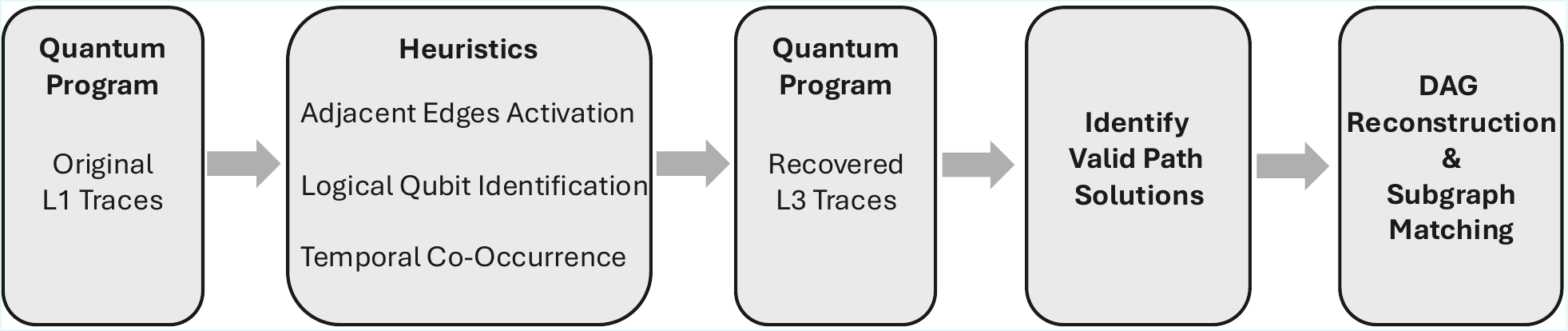}
  \caption{\small TraceQ workflow for trace analysis.}
  \label{fig_trace_analysis_workflow}
\end{figure*}

In fault-tolerant quantum computers, the quantum circuit is mapped onto a 2D qubit lattice, where each two-qubit gate induces a routing path between control and target qubits. These physical paths can be recorded or simulated as discrete time-indexed matrices encoding the activity of each patch on the grid. Such \textit{access traces} can offer a powerful window into the structure and scheduling information of the original quantum circuit, but they also introduce significant challenges due to parallelism, density and ambiguity.

\subsection{Levels of Access Traces}

We describe our {\em access traces} in increasing levels of information; that is, a Level-1 trace has the least amount of information. We define levels of traces as:
\begin{enumerate}
  \item \textbf{Level-1 (L1)}: only contains free/busy status of each error corrected patch, i.e. 1 bit of information per patch for each time step.
  \item \textbf{Level-2 (L2)}: Level 1 \textit{plus} the free/busy status of each boundary between logical qubits; each patch has 4 sides or boundaries, so in total there are 5 bits of information per patch for each timestep.
  \item \textbf{Level-3 (L3)}: Level 2 \textit{plus} identification if a patch corresponds to a logical qubit or a mapping/routing region and if this trace entry has multiple valid paths to connect qubit endpoints. 
  In total there are 7 bits of information per patch for each timestep.
\end{enumerate}

\subsection{Access Trace Formats}

Each trace is a sequence of 2D matrices, each representing a timestep of gate activity of the logical qubits across a grid of tiles. In the L3 trace format, each matrix entry contains a 7-bit encoding that captures both the functional role of the tile and its directional connectivity. Specifically, the second and the third most significant bits represent the qubit’s role, \texttt{10} for control qubit, \texttt{11} for target qubit, and \texttt{01} for connection tile, while the remaining four bits encode edge activity toward neighboring tiles in the north, south, west, and east directions. The most significant bit represents if a trace entry is ambigious or not (i.e., whether there are multiple valid, non-overlapping paths to connect the qubit endpoints). This format provides precise reconstruction of qubit roles and paths for each gate which executes, but assumes access to detailed, low-level spatial execution data. Level-2 traces extend the basic patch activity of Level-1 by providing per-patch boundary state~information. 

While L3 traces contain rich structural detail, a more realistic and constrained setting is one where only~Level-1 (L1) trace information is observable. In the L1 format, each timestep $t \in {1, \dots, T}$ yields a matrix $M_t \in {0,1}^{d \times d}$, where:
\begin{itemize}
\item $M_t[r, c] = 1$ indicates that patch $(r, c)$ is active,
\item $M_t[r, c] = 0$ indicates that patch $(r, c)$ is idle.
\end{itemize}

The L1 format omits all qubit roles and path information, and thus no distinction is made between control and target qubits, nor are routing directions preserved. This representation models what may be available in practice in the current fault-tolerant protocols and makes minimal assumptions about what information is observable from the control stack.

\subsection{Reconstructing Program DAG from Access Traces}

The core problem addressed in this work is the reconstruction of the logical dataflow of a quantum program from such trace sequences. The ultimate goal is to recover all two-qubit gates and their temporal dependencies, even in the presence of parallelism, ambiguity, and incomplete information of the traces. Formally, let $\mathcal{T} = { M_1, M_2, \dots, M_T }$ be the input sequence of $d \times d$ spatial matrices. In the L3 format, each nonzero entry $M_t[r, c]$ may encode a specific qubit role, directionality, or reserved patterns to handle later ambiguous traces. In contrast, in the L1 format, each matrix is binary-valued, and the only observable information is patch-level activity (only if it is active or not).

Reconstructing a directed acyclic graph (DAG) representing the program’s gate operations from these L1 traces is the main challenge that we address in this work. Our methods focus on this setting, and all evaluations in this work are conducted using L1 traces as input. We further focus on reconstruction of DAG with only two-qubit gate operations.

\section{TraceQ Framework and Software}
\label{software_implementation}

To reconstruct the dataflow of quantum programs from \textit{access} traces, we developed TraceQ, which is a trace-based software framework that analyzes the spatial-temporal activity of logical qubits on a 2D grid layout as captured in the activity traces. The framework is written in Python and leverages mainly Qiskit~\cite{javadiabhari2024quantumcomputingqiskit}, OpenQASM~\cite{cross2017openquantumassemblylanguage} and NetworkX~\cite{osti_960616}. A schematic of the TraceQ end-to-end workflow is shown in Figure~\ref{fig_trace_analysis_workflow}.

\subsection{Trace Generation and Encoding}

In order to analyze how L1 traces can be used to recover program dataflow, we need to first generate such traces for testing.
Our trace generation pipeline begins by parsing a quantum program in OpenQASM format and converting it to a QuantumCircuit object via Qiskit. Each logical qubit is mapped to a unique patch location in a configurable layout. To determine gate scheduling, we analyze the circuit’s DAG structure and extract all levels of parallelizable two-qubit operations, mainly CNOT and CZ gates.
The specific two-qubit gate type does not matter for the trace encoding.

Each group of parallel gates is routed using a custom A*-based pathfinding algorithm on a 2D grid graph, where logical qubit locations serve as source and destination nodes. The gates are scheduled using an as-many-as-possible policy using a greedy heuristic algorithm. Paths are constructed to avoid physical overlaps and inactive qubit positions, and each patch visited during the routing process is annotated with a binary encoding that captures only whether the patch is active or not.
The resulting binary matrices are saved as the original L1 traces, indexed by timestep. 

The result is a minimal, binary abstraction of the original trace that more closely reflects what a control stack might expose. Our main goal is to reconstruct the dataflow of quantum programs from these~traces.

\subsection{Heuristics Recovery of L3 Traces from L1}
\label{heuristics_l3_to_l1}

\begin{figure}[t]
    \centering
    \includegraphics[width=0.855\linewidth,trim={0cm 0cm 0cm 0cm},clip]{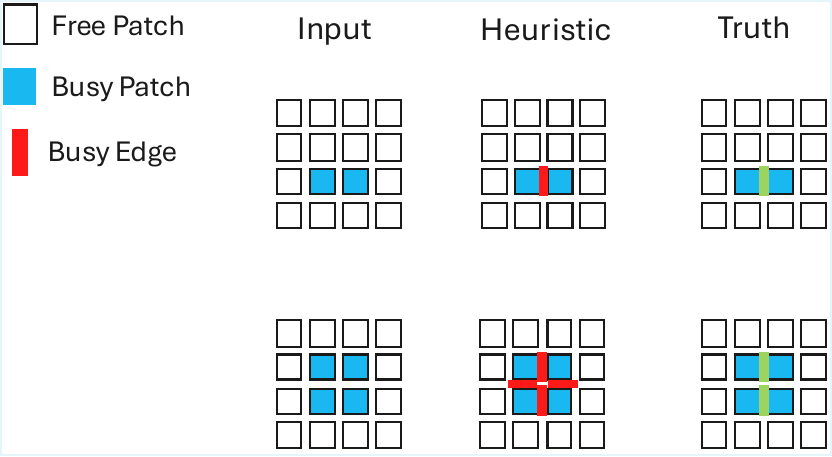}
    \caption{\small Given an L1 trace, we annotate each edge between patches as 1 if both patches which are adjacent to the edge are active and 0 otherwise. This heuristic is too generous and overlabels edges due to ambiguity. With light green color we show the actual busy edges.}
    \label{fig_heuristic_edges}
\end{figure}

\begin{figure}[t]
    \centering
    \includegraphics[width=0.855\linewidth,trim={0cm 0cm 0cm 0cm},clip]{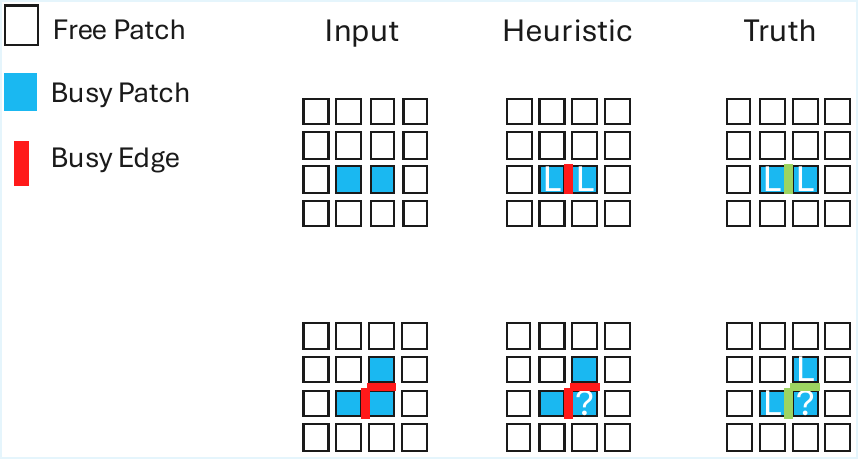}
    \caption{\small Given an L1 trace, we attempt to find patches which touch only one edge (node degree=1). This might not necessarily find all logical qubits, but will never mislabel them. We indicate an "L", if a logical qubit is identified. With light green color we show the actual busy edges.}
    \label{fig_heuristic_qubits}
\end{figure}

\begin{figure}[t]
    \centering
    \includegraphics[width=0.855\linewidth,trim={0cm 0cm 0cm 0cm},clip]{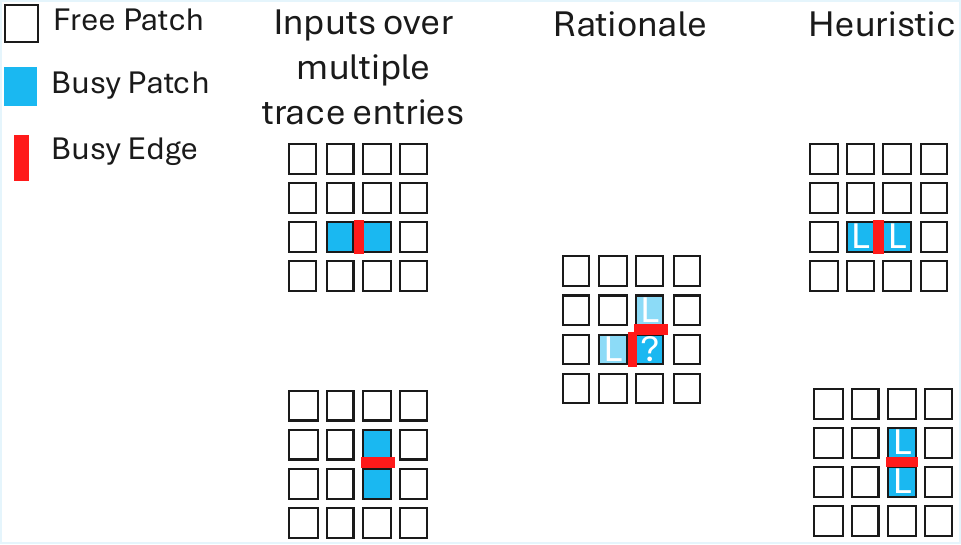}
    \caption{\small Given an L1 trace, we utilize temporal co-occurrence knowledge from other trace entries, and thus we are able to infer logical qubit locations on the grid. We indicate an "L", if a logical qubit is identified.}
    \label{fig_heuristic_temporal_co_occurence}
\end{figure}

To recover fine-grained control and connectivity information from the least informative L1 trace representation, we implement a set of heuristics that reconstruct L3 traces from L1 traces. The L3 traces capture the full dataflow, which is what we want to be able to reconstruct, but also by having generated L3 traces before L1 we can use them as reference. The L1 format retains only free/busy activity at the patch level, that is, which grid patches are active, but omits explicit role (control/target) and directionality information. Our recovery process proceeds in three stages, each having an associated heuristic:

\subsubsection{Heuristic 1: Adjacent Edges Activation}

Given an L1 trace entry, which is just free/busy patch knowledge, we mark an edge as active, if we observe two adjacent patches to be active. This heuristic can significantly over-label edges due to the inherent ambiguity of L1 trace. For instance, there might be a case where two patches are next to one another and both are active, however, they do not need to be interacting with each other. Although this heuristic might be generous, it is essential in order to formulate the (possible) paths that connect the qubit endpoints, which we identify next. Figure~\ref{fig_heuristic_edges} depicts how the edges between two patches are annotated as active.

\subsubsection{Heuristic 2: Logical Qubit Identification}

Next stage is to identify patches containing logical qubits. We  identify all patches likely to correspond to logical qubits using local connectivity criteria. A patch is considered a candidate logical qubit if it is active and has exactly one active neighbor in the L1 trace. This is implemented using a $3\times3$ convolution kernel to compute neighbor sums, followed by masking with the activity map. This heuristic might not identify all logical qubit locations, as the borders of some patches might not be informative, as we show in Figure~\ref{fig_heuristic_qubits}, but it will not falsely identify incorrect ones. To identify the remaining logical qubits, we leverage temporal co-occurrence as we see next.

\subsubsection{Heuristic 3: Temporal Co-Occurrence}

Figure~\ref{fig_heuristic_temporal_co_occurence} shows how we leverage temporal co-occurrence information from other trace entries to be able to recover more logical qubit locations on the layout.
After extracting as many logical qubit locations as possible with Heuristic $1$ and $2$, we sieve the traces again. We propagate this temporal co-occurrence knowledge to trace entries where the qubit locations were not initially identified. The goal of this heuristic is to inadvertently catch the routing space, and potentially infer all the logical qubit locations in the layout. Next, the L1 trace is partitioned into disjoint connected components. Within each component, we distinguish between known paths, i.e., the unambiguous pairwise connections between logical qubits, and ambiguous regions that cannot be reliably resolved. For known paths, we construct a constricted logical subgraph: a graph over logical qubits where an edge exists between two qubits if a path connects them without passing through other logical qubits. We then greedily decompose this graph using breadth-first search (BFS), starting from degree-1 nodes, to extract non-overlapping paths. Remaining subgraphs with unresolved connectivity are treated as unknown, and thus these are marked as ambiguous regions. After heuristic $3$ is completed, we are able to move from the lossy, free/busy L1 trace level, to the most informative L3 trace level.

\subsection{Identifying Potential Path Solutions in an Ambiguous Recovered L3 Trace}

\begin{figure}[t]
    \centering
    \includegraphics[width=1\linewidth,trim={0cm 0cm 0cm 0cm},clip]{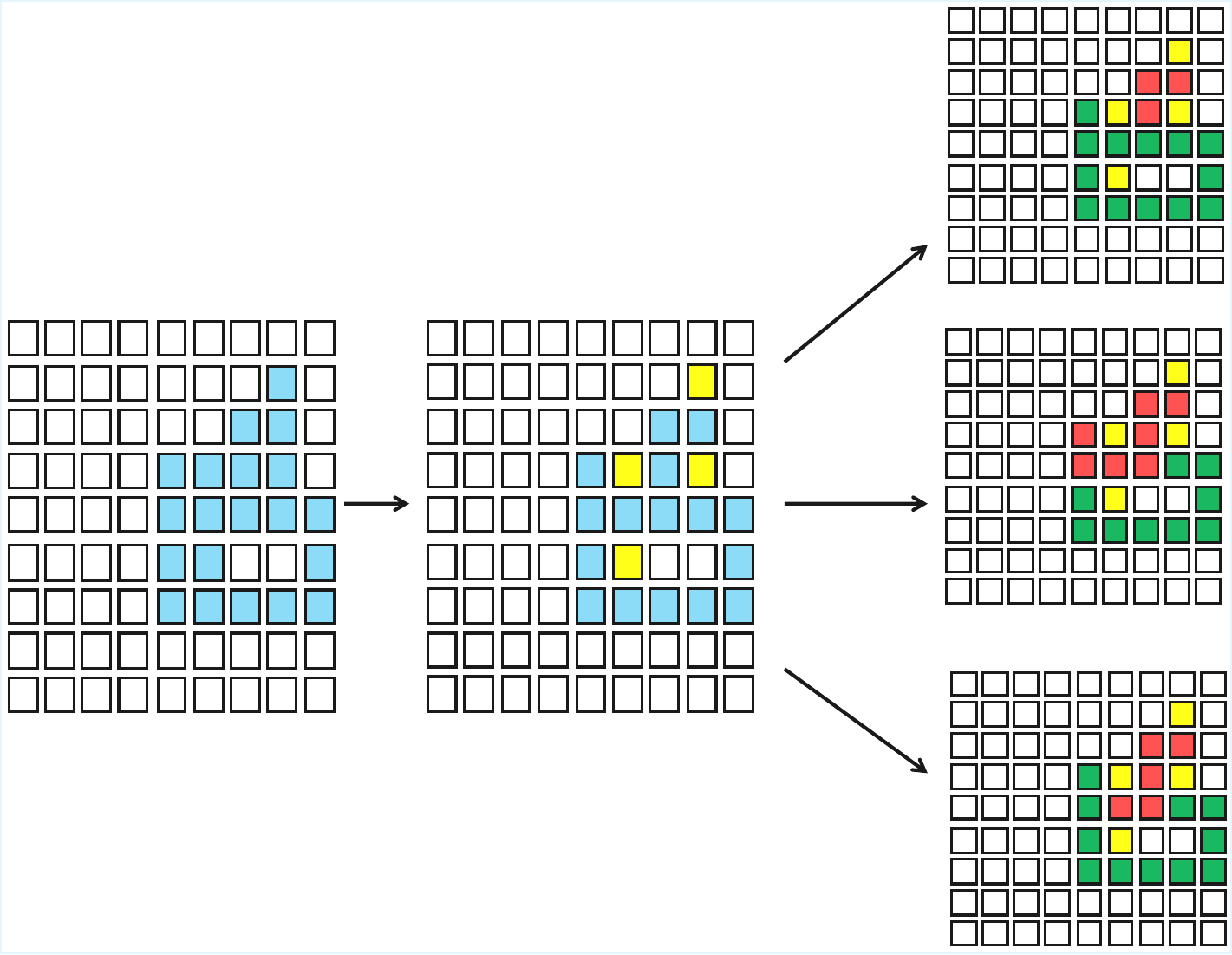}
    \caption{\small Example of a resolution of an ambiguous trace entry. After recovering L3 traces from original L1 level and identifying the logical qubit endpoints (yellow) from L1 trace, through a DFS algorithm we are able to find all valid, non-overlapping possible path solutions (with red and green). All active patches have to be utilized by the DFS to yield a valid solution.}
    \label{fig_dfs_possible solutions}
\end{figure}

We now have converted L1 trace to a reconstructed L3 traces that has some ambiguity. For each ambiguous L3 trace entry, we initialize a new L3 matrix by labeling logical patches as logical qubit endpoints, which correspond to unknown paths with a reserved encoding as $0b1110000$ and wire patches as generic connections between two logical qubit endpoints, as $0b1010000$. In order to find all the valid routing paths between the qubit endpoints we need to apply a traversal algorithm, such as Depth-First Search (DFS) algorithm. Figure~\ref{fig_dfs_possible solutions} shows the possible (valid) path solutions, after running a DFS algorithm with backtracking logic to identify valid, non-overlapping paths. The DFS algorithm is mainly driven by the wire patches, that is, $0b1010000$ elements. In certain circumstances, though, we may have cases where two logical qubits are adjacent and thus no intermediate wire patch is required in between. The paths should not cross other inactive logical qubits, nor crossing other paths from the same solution. In order for a path to be valid, it also has to utilize all active patches at that \textit{access} trace entry. Known paths are further annotated using a custom convolution-based filter that infers directional edge activity between adjacent patches (N/S/W/E) and merges this into the $7$-bit matrix format. Patches belonging to unresolved components are marked with the MSB=$1$ in a $7$-bit trace format to indicate ambiguity in both role and routing. The Most Significant Bit (MSB) represents whether this is an ambiguous trace at this timestep if MSB is set, or refers to an unambiguous path~otherwise.

\subsection{DAG Reconstruction and Subgraph Detection}
Finally, to reconstruct the full logical dataflow, we instantiate a directed acyclic graph (DAG) where nodes represent two-qubit gates and edges reflect temporal or data dependencies. For unambiguous steps, gates are deterministically added. For ambiguous steps, we insert a separate gate node for each valid configuration and link all dependencies accordingly. Per-qubit histories are maintained to determine lookahead constraints when inserting edges, ensuring correctness in gate ordering. The result is an augmented, ambiguity-aware program~DAG.

We further enable subroutine detection by applying exact subgraph isomorphism between reconstructed program DAGs and known subcircuit DAGs. To implement this, we use the VF3 algorithm~\cite{carletti2017introducing}, which is the state-of-the-art subgraph matching algorithm to use for huge and dense graphs, and also efficient in time and memory, in comparison with its predecessors. This allows us to confirm the presence of known subroutine patterns (e.g., adder, QFT)
embedded within a larger program execution trace. All \textit{access} trace entries and graphs are exported in JSON files for downstream analysis. We will make the code of our framework publicly available for~reproducibility.

\section{Evaluation}
\label{evaluation}

We now proceed to evaluate the TraceQ framework and show its ability to recover the dataflow and identify programs and subroutines from the recovered traces.

\begin{table}[t]
  \centering
  \footnotesize %
  \setlength{\tabcolsep}{4pt} %
    \scalebox{0.69}{
  \begin{tabular}{@{} l c c ccc @{}} 
    \toprule
    \textbf{Synthetic} & \textbf{Instances} & \textbf{\# Qubits} & \multicolumn{3}{c}{\textbf{Avg. DAG Nodes}} \\
    \cmidrule(l){4-6}
                       &                    &                   & \textbf{Compact} & \textbf{Sparse} & \textbf{Interm.} \\
    \midrule
    random\_mix\_1     & qft\_5 + t\_npe\_4 + add\_3         & 29  & 207 & 376 & 206 \\
    random\_mix\_2     & qft\_6 + t\_npe\_3 + add\_4         & 26  & 153 & 205 & 151 \\ 
    random\_mix\_3     & qft\_4 + t\_npe\_4 + add\_5         & 34  & 225 & 395 & 236 \\ 
    random\_mix\_4     & qft\_11 + t\_npe\_5 + add\_7        & 56  & 439 & 1106 & 446 \\ 
    random\_mix\_5     & t\_npe\_3 + add\_9                  & 35  & 173 & 202 & 173 \\ 
    random\_mix\_6     & qft\_13 + add\_10                   & 42  & 265 & 269 & 263 \\ 
    random\_mix\_7     & qft\_13 + t\_npe\_5                 & 38  & 413 & 985 & 423 \\
    random\_mix\_8     & add\_11 + t\_npe\_6                 & 68  & 496 & 1492 & 519 \\
    random\_mix\_9     & add\_3 + qft\_4                     & 12  & 36 & 36 & 34 \\ 
    random\_mix\_10    & add\_4 + qft\_4                     & 15  & 48 & 48 & 47 \\ 
    random\_mix\_11    & prod\_3 + outofplace\_add\_3 + qft\_4 & 32  & 173 & 205 & 173 \\ 
    random\_mix\_12    & prod\_3 + cuccaro\_4 + qft\_4       & 32  & 152 & 176 & 152 \\ 
    random\_mix\_13    & t\_npe\_3 + cuccaro\_7              & 25  & 112 & 204 & 113 \\
    random\_mix\_14    & outofplace\_add\_5 + qft\_10 + add\_5 & 40  & 199 & 297 & 195 \\
    random\_mix\_15    & add\_4 + outofplace\_add\_4 + cuccaro\_4 & 34 & 105 & 230 & 102 \\
    random\_mix\_16    & t\_npe\_3 + qft\_5 + cuccaro\_9     & 34  & 142 & 311 & 147 \\ 
    random\_mix\_17    & cuccaro\_12 + qft\_4 + t\_npe\_3    & 39  & 152 & 424 & 157 \\ 
    random\_mix\_18    & add\_12 + t\_npe\_3 + qft\_6        & 50  & 238 & 331 & 243 \\ 
    random\_mix\_19    & t\_npe\_4 + add\_12 + qft\_5        & 56  & 307 & 469 & 312 \\
    random\_mix\_20    & qft\_9 + cuccaro\_13 + adder\_12    & 72  & 270 & 482 & 270 \\
    \bottomrule
  \end{tabular}
    }
  \caption{\small Composition, qubit count, and average DAG node counts per layout for each synthetic benchmark.}
  \label{tab_synthetic-benchmark}
\end{table}

\subsection{Subroutine Benchmark Selection}
\label{subroutine_selection}

While there exist benchmarks for the NISQ era \cite{tomesh2022supermarq,li2023qasmbench}, fault-tolerant quantum programs will vary significantly. For example, fault-tolerant quantum chemistry algorithms require large-scale arithmetic circuits that are inaccessible to NISQ devices \cite{king2025quantum, von2021quantum}. We synthesize our own benchmarks to represent key quantum subroutines. We produce variable size implementations of adders, a product, the Quantum Fourier Transform (QFT), and a sample Trotterization circuit for various steps without phase estimation. These subroutines are representative of components in larger algorithms (e.g., Shor’s algorithm uses modular adders and QFT and Trotterization is central to chemistry simulations). The quantum subroutine circuits were generated with pyLIQTR~\cite{obenland_2025_15492111}, a library designed for constructing quantum circuits based on quantum algorithms.

\begin{figure}[t]
\vspace{0.08cm} 
  \centering
  \includegraphics[width=1\linewidth]{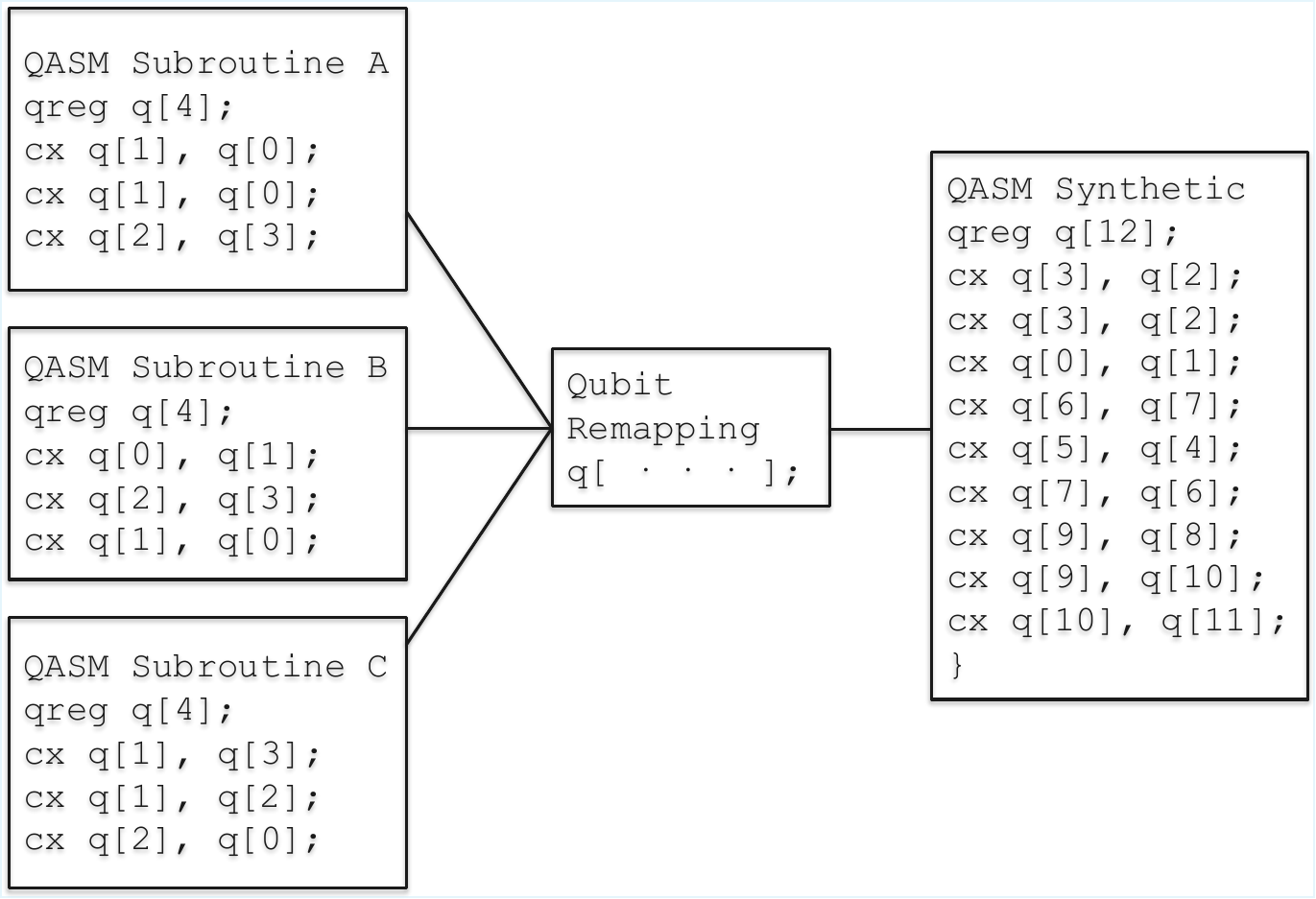}
  \caption{\small Example of synthetic benchmark generation. We get 30 different random samples of the qubit labelings for each generated synthetic benchmark.}
  \label{fig_synthetic_bench_gen}
\end{figure}

\subsection{Synthetic Benchmark Generation}
\label{synethetic_benchmarks}

Table~\ref{tab_synthetic-benchmark} summarizes the $20$ synthetic benchmarks were used for benchmarking our~work.
To construct scalable and structurally diverse quantum benchmarks, we implement a function that creates synthetic quantum circuits that are
composed of useful and common quantum building blocks, such as adders and QFTs, into a single unified QASM file. 
This is achieved through random qubit index remapping and concatenation of gate instructions, enabling the formation of composite workloads that create complex circuits while preserving internal gate structures of known subroutines. In Figure~\ref{fig_synthetic_bench_gen}, a schematic shows how subroutines are being remapped and merged to generate a larger synthetic~program. By validating TraceQ on combinations of these subroutines, we simulate realistic workload patterns.

Each input QASM file is parsed independently, and its qreg declarations are identified and tracked. To prevent naming collisions during integration, logical qubit indices are globally renumbered across all subcircuits. Within each subroutine, we randomly permute the local qubit indices before inserting the instruction body into the composite program. This randomization introduces structural variability in the resulting program DAG.
\begin{figure}[t]
  \centering
  \includegraphics[width=1\linewidth]{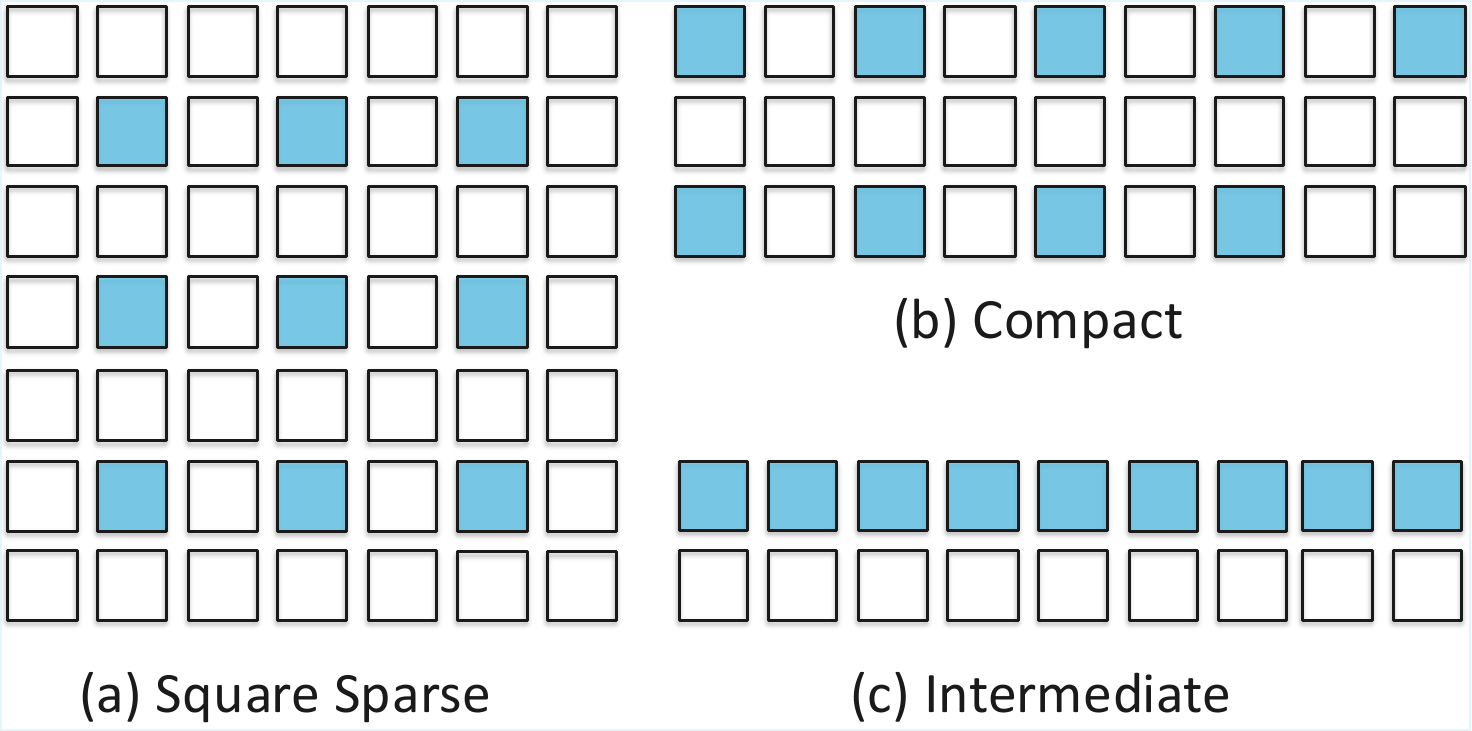}
  \caption{\small Given a $9$-qubit circuit, we present three logical‐qubit layouts used in this work: Square Sparse (left), Compact (top-right), and Intermediate block (bottom-right).}
  \label{fig_three_layouts}
\end{figure}

\begin{figure*}[t]
    \centering
    \includegraphics[width=1\linewidth,trim={0cm 0cm 0cm 0cm},clip]{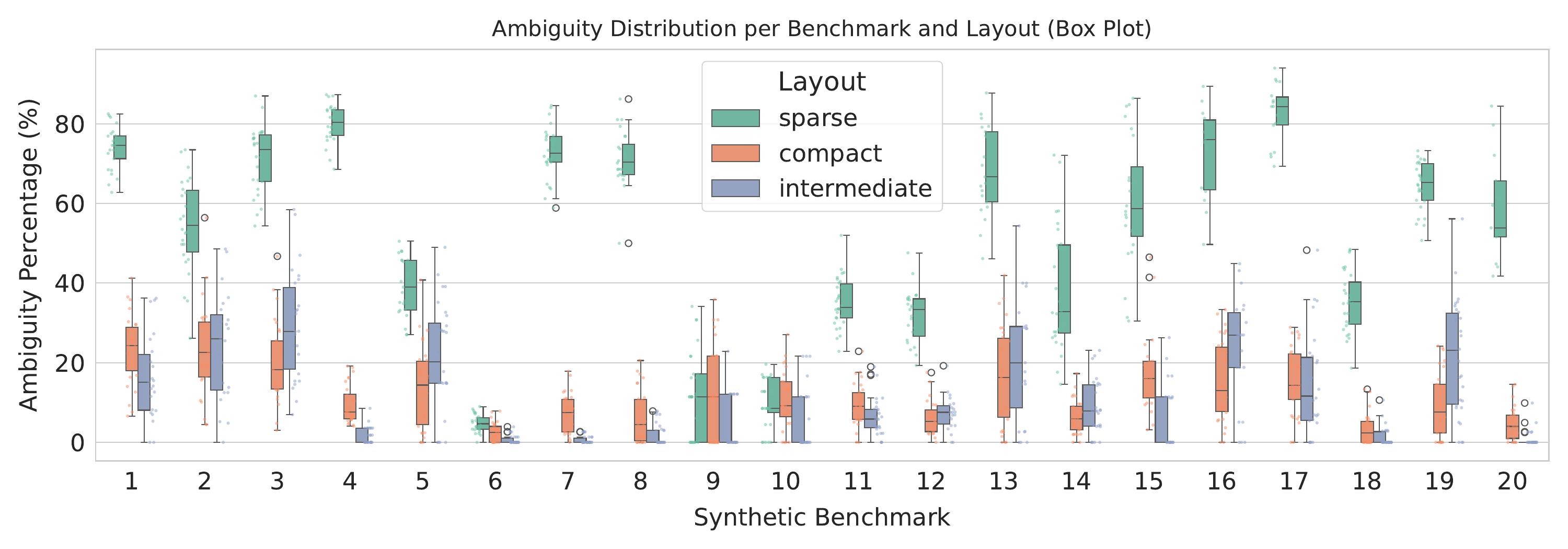}
    \caption{\small Box plot showing the distribution of ambiguity percentages across $20$ synthetic benchmark circuits for each layout type. Each box summarizes the variation over $30$ perturbations. The ambiguity percentage quantifies the fraction of trace entries labeled as ambiguous in each perturbation.}
    \label{fig_box_plot_ambiguity}
\end{figure*}

The output is a valid QASM program that declares a unified qubit register encompassing all merged instructions. The resulting benchmark can be used to evaluate the scalability of trace reconstruction, subgraph detection, and ultimately, how robust our framework is under unknown mapping strategies. We provide $20$ random synthetic benchmarks, and for each one of them we provide $30$ different random perturbations of the qubit labelings, so in total $20 \times 30$ = $600$ different~cases.

\subsection{Layout Architectures}
\label{layouts}
We design our approach to support arbitrary architectures.
Figure~\ref{fig_three_layouts} depicts three standard layouts proposed in the literature~\cite{horsman2012surface, litinski2022active, watkins2024high, 10.1145/3720416} that we used to evaluate our framework, which we call the Square Sparse, Compact and Intermediate layouts.
Square Sparse (Fig.~\ref{fig_three_layouts}a) and Compact (Fig.~\ref{fig_three_layouts}b) architectures represent opposite ends of the space-time spectrum, with the Square Sparse architecture using many more qubits to allow routing more gates in parallel.
Intermediate (Fig.~\ref{fig_three_layouts}c) layout can be considered as an improved version (in terms of time) of the Compact layout as it can store all qubits in one row instead of two.
The exploration of efficient floorplans is crucial for solving large
problems with a shorter execution time and fewer qubits.

\begin{figure*}[t]
    \centering
    \includegraphics[width=1\linewidth,trim={0cm 0cm 0cm 0cm},clip]{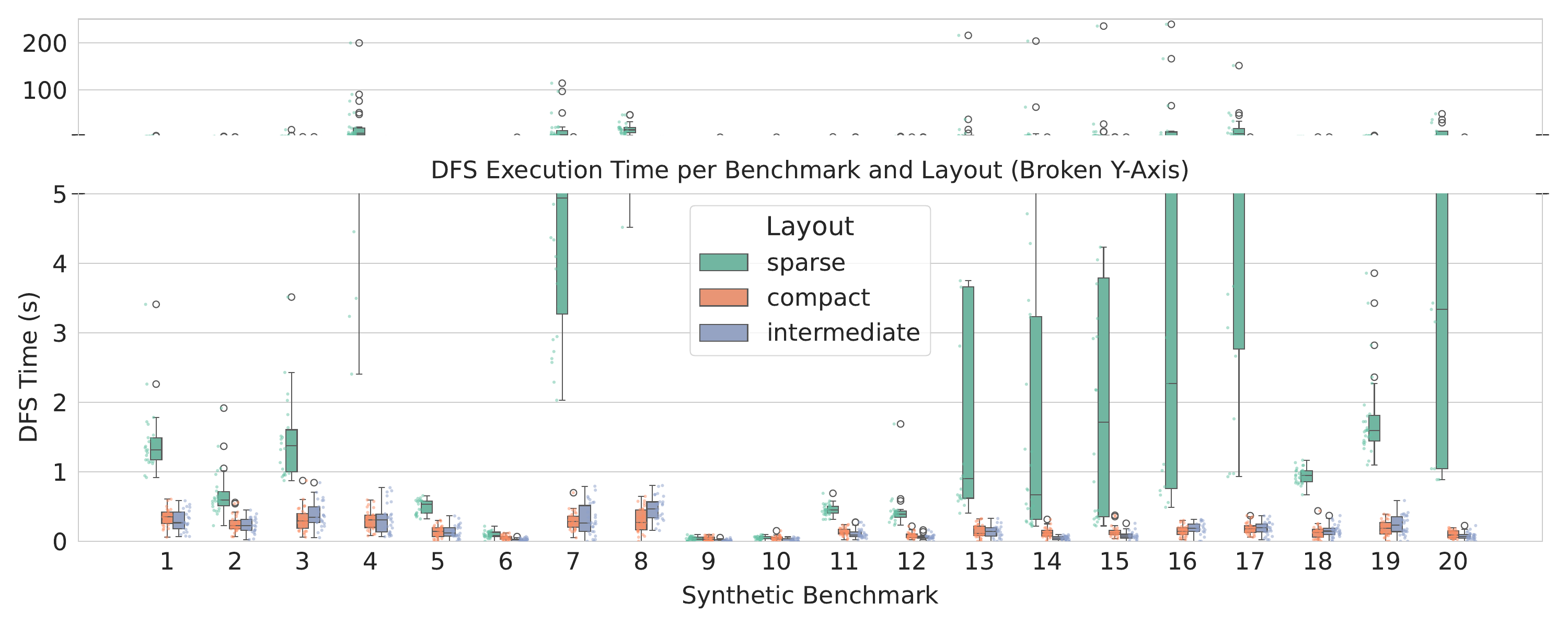}
    \caption{\small Box plot of DFS execution times across 20 synthetic benchmarks and three layout strategies. The y-axis is broken at $5$ seconds to reveal both the main distribution of low-latency executions and a few significantly longer outliers. Each box summarizes $30$ perturbations per benchmark-layout pair, excluding unsuccessful runs.}
    \label{fig_boxplot_dfs_runtime}
\end{figure*}

\begin{figure}[t]
    \centering
    \includegraphics[width=0.8\linewidth,trim={0cm 0cm 0cm 0cm},clip]{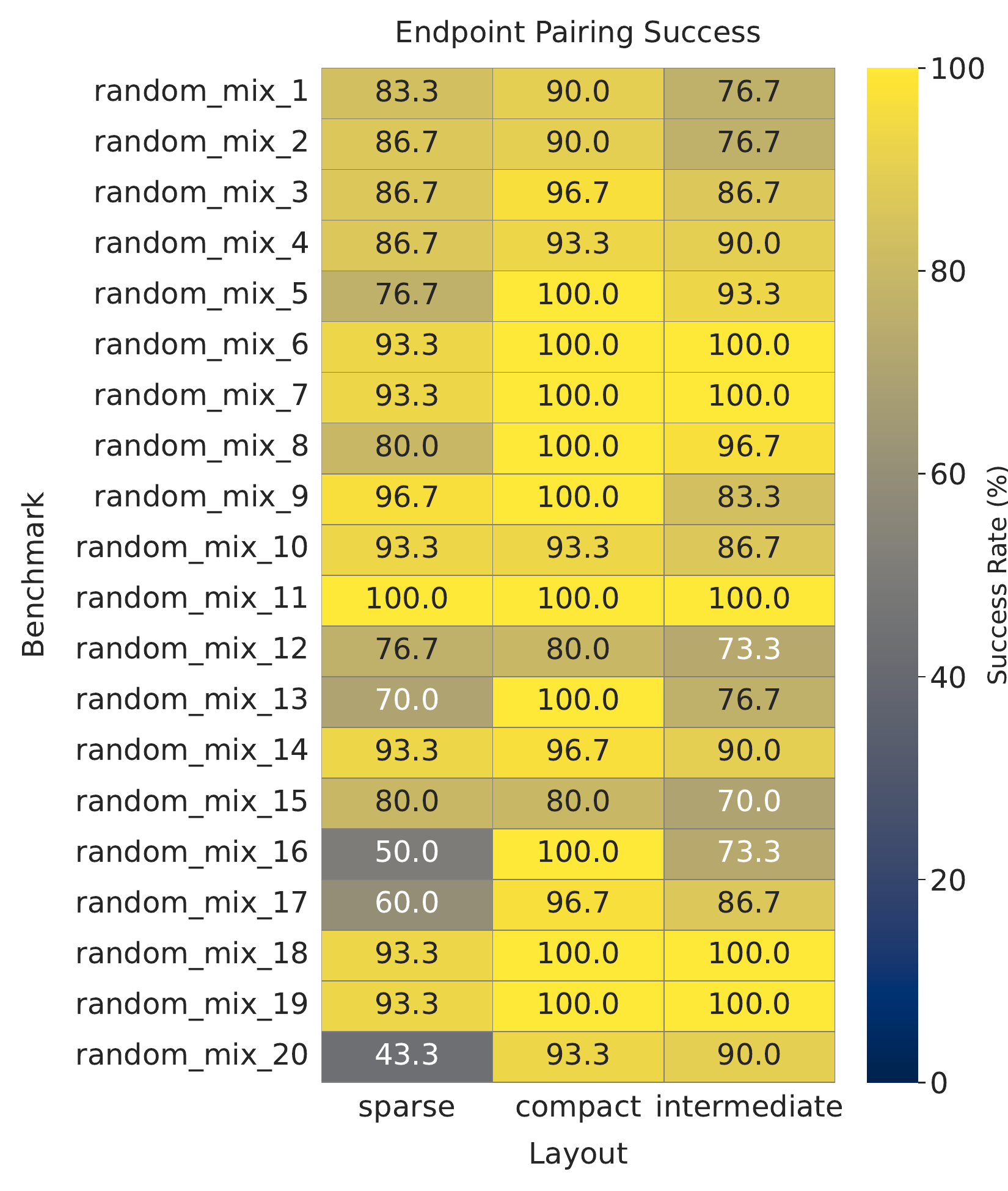}
    \caption{\small Heatmap of success rates for $20$ synthetic quantum benchmarks across three layout architecture after post-DFS processing. Each cell indicates the percentage of perturbation instances (out of $30$) that successfully completed without an odd-ambiguity error. Colormap reflects success rate from $0$\% (dark) to $100$\% (bright yellow).}
    \label{fig_post_dfs_success_rate_v5}
\end{figure}

\begin{figure}[t]
  \centering
  
  \begin{subfigure}{\linewidth}
    \centering
    \includegraphics[width=0.65\linewidth]{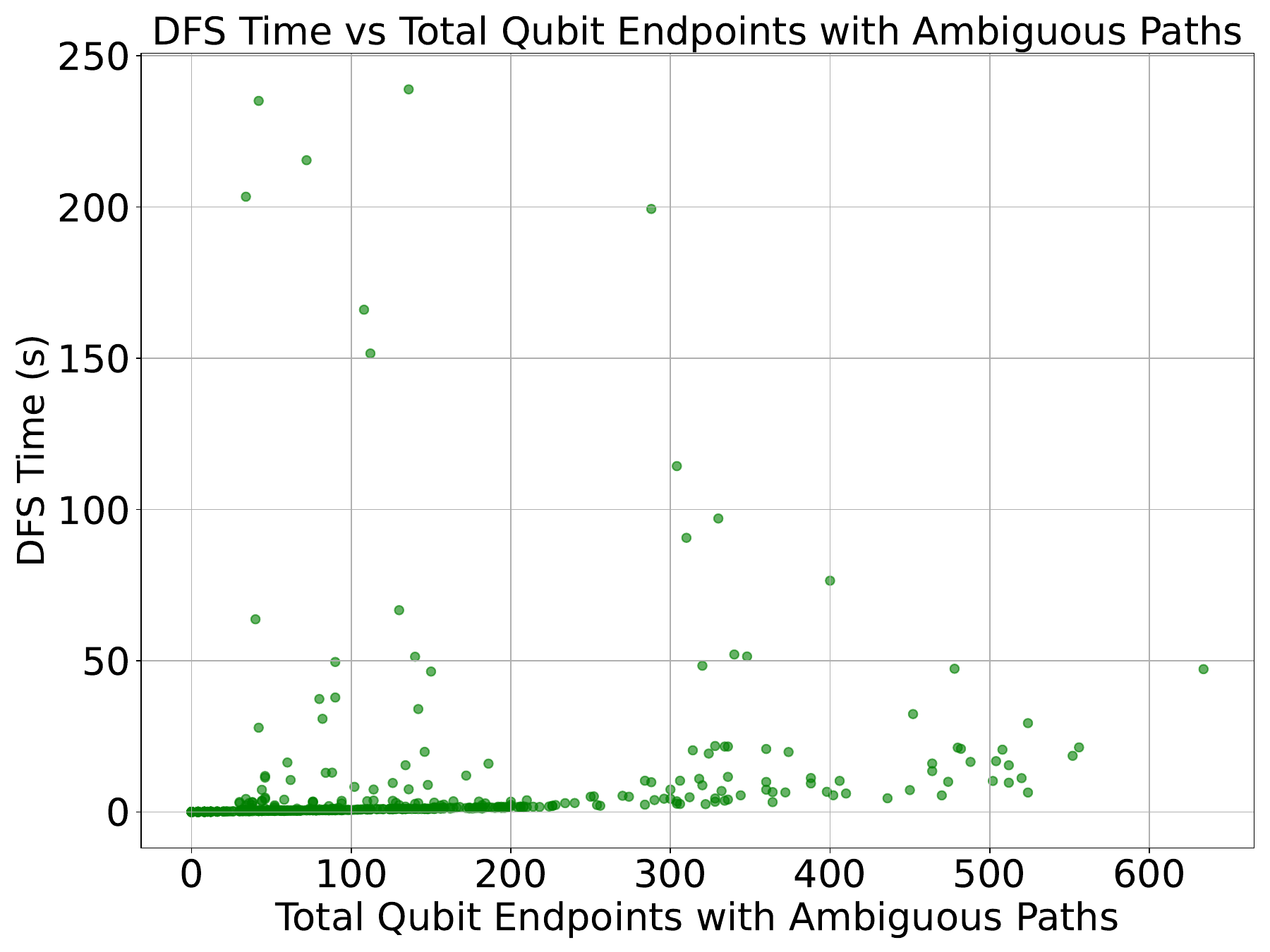}
    \caption{Square Sparse}
    \label{fig_vs_unknown_cells_sparse}
  \end{subfigure}
  
  \vspace{1em}
  
  \begin{subfigure}{\linewidth}
    \centering
    \includegraphics[width=0.65\linewidth]{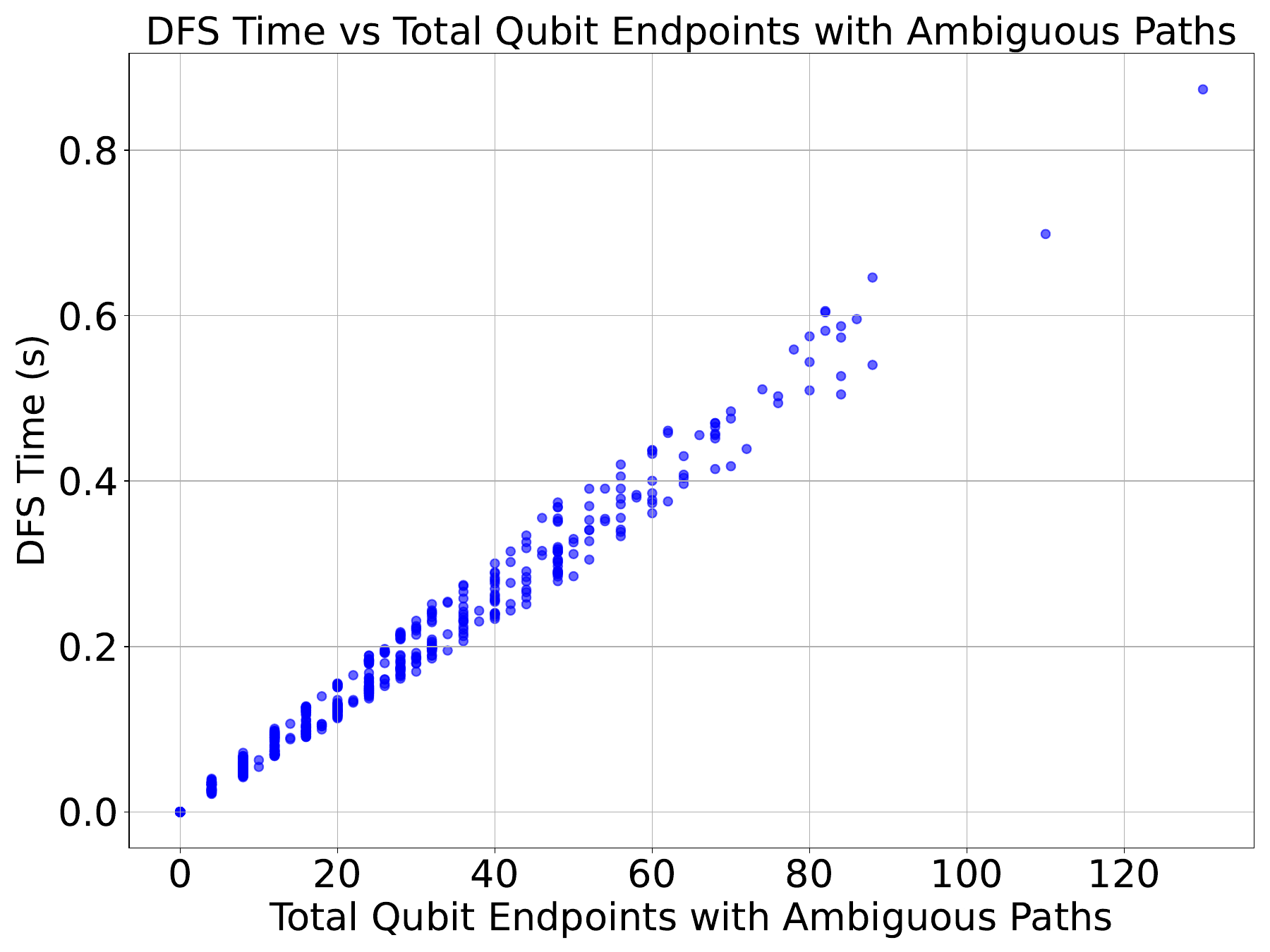}
    \caption{Compact}
    \label{fig_vs_unknown_cells_compact}
  \end{subfigure}
  
  \vspace{1em}
  
  \begin{subfigure}{\linewidth}
    \centering
    \includegraphics[width=0.65\linewidth]{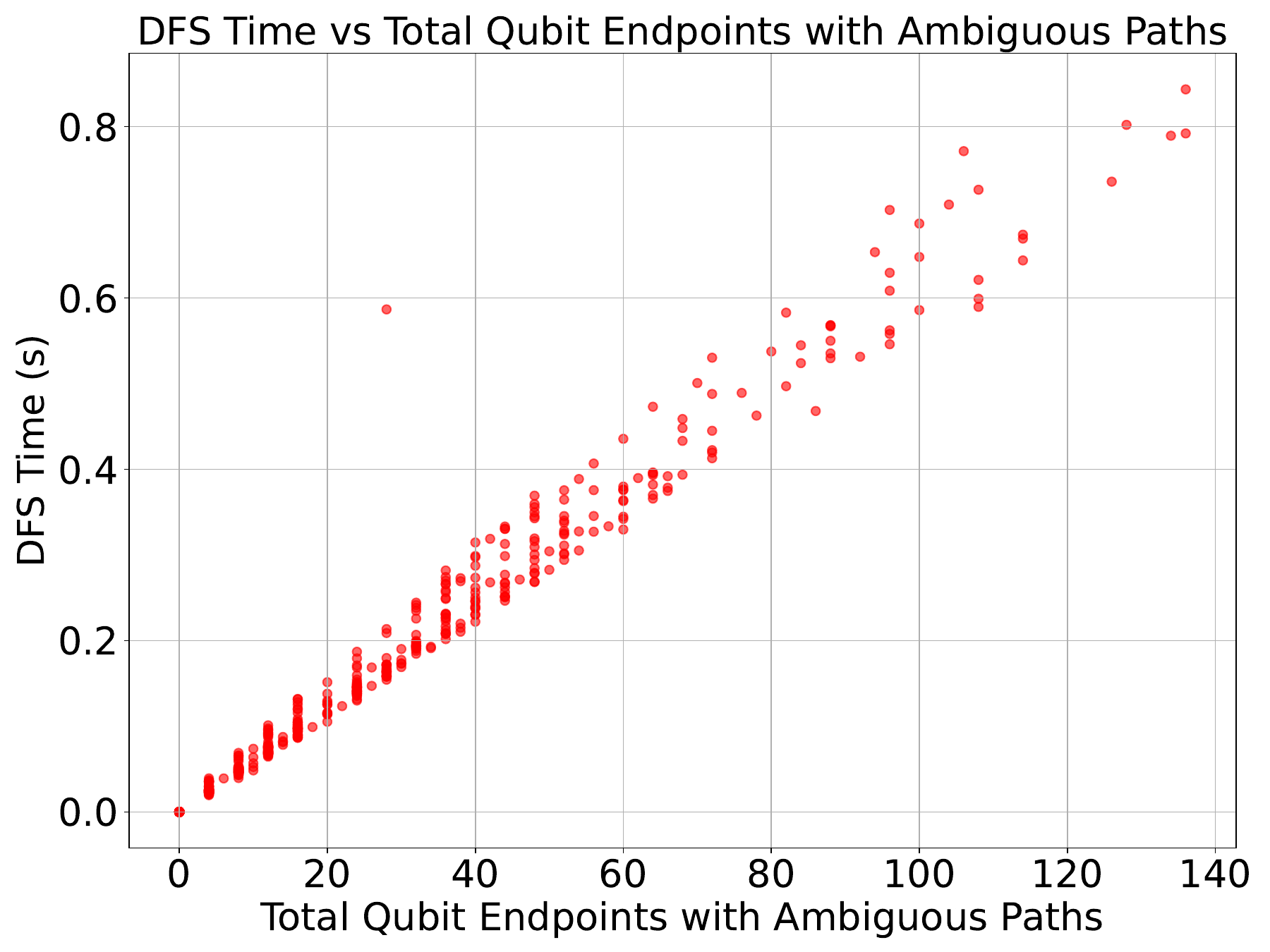}
    \caption{Intermediate}
    \label{fig_vs_unknown_cells_intermediate}
  \end{subfigure}

  \caption{\small Relationship between DFS execution time and the number of qubit endpoints with ambiguous routing paths for each layout. Each point corresponds to a single perturbation instance of a synthetic benchmark. Layouts with more spatial degrees of freedom (e.g., Square Sparse) exhibit higher ambiguity levels and longer DFS resolution times, suggesting a correlation between layout routing complexity and runtime.}
  \label{fig_dfs_layouts}
\end{figure}

\subsection{Results}
\label{evaluation_results}

\begin{figure}[t]
  \centering

  \begin{subfigure}{\linewidth}
    \includegraphics[width=1\linewidth]{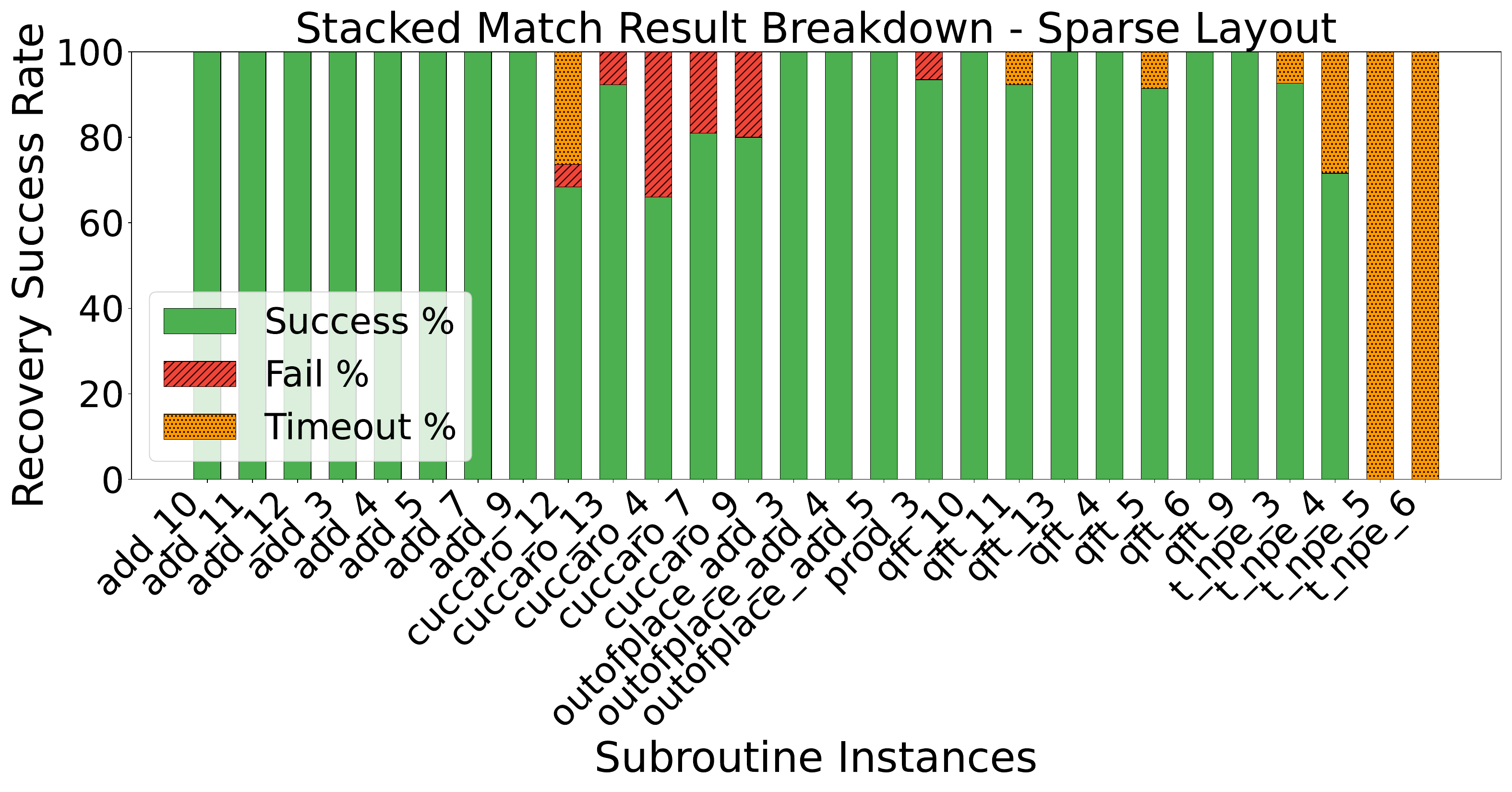}
    \caption{\small Square Sparse}
    \label{fig_layout_sparse_results}
  \end{subfigure}

  \begin{subfigure}{\linewidth}
    \includegraphics[width=1\linewidth]{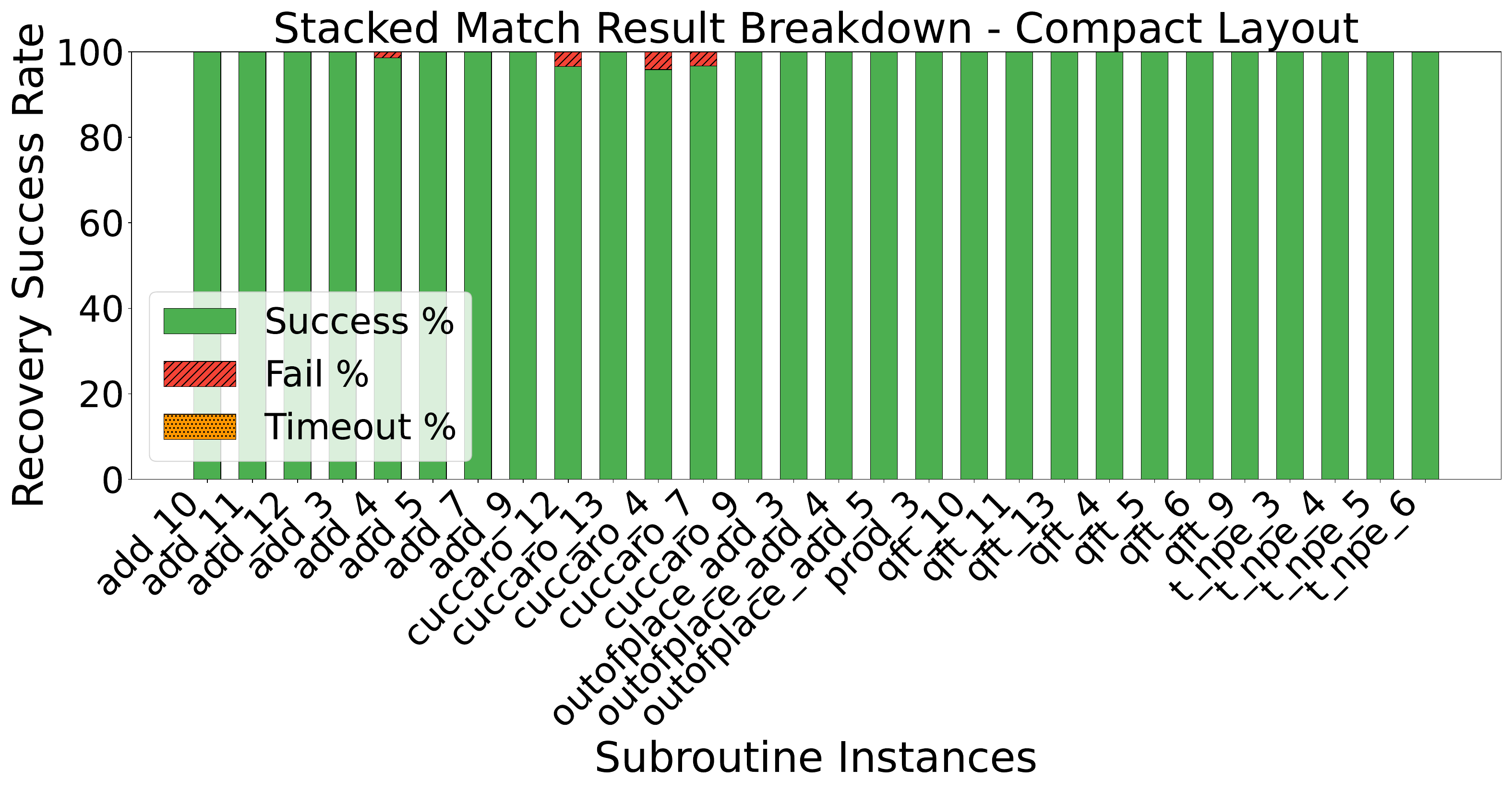}
    \caption{\small Compact}
    \label{fig_layout_compact_results}
  \end{subfigure}

  \begin{subfigure}{\linewidth}
    \includegraphics[width=1\linewidth]{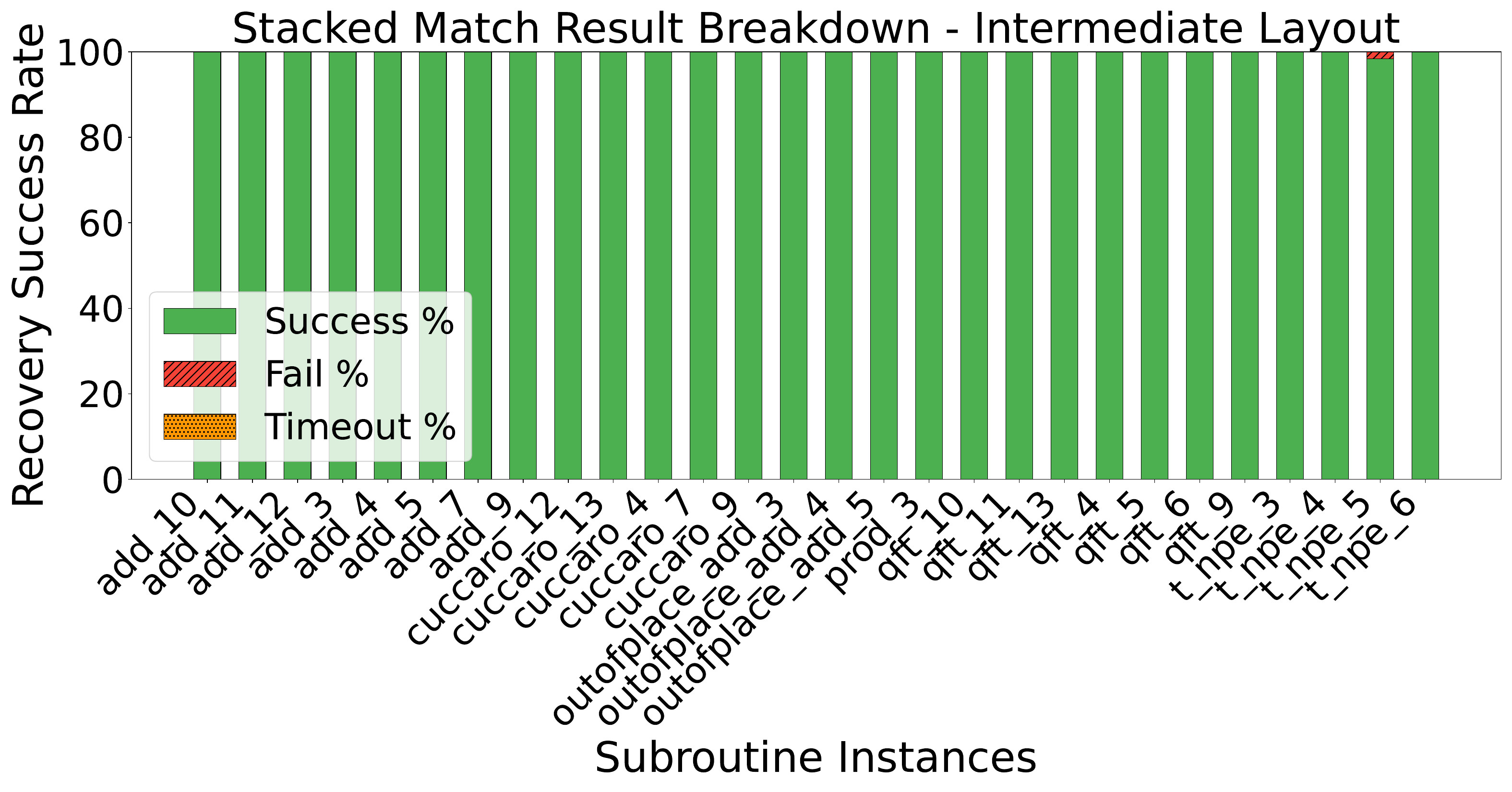}
    \caption{\small Intermediate}
    \label{fig_layout_intermediate_results}
  \end{subfigure}

  \caption{\small Stacked success (green), failure (red), and timeout (orange) percentages per subroutine, aggregated over all benchmarks, across three layouts for the subgraph matching process. Each subroutine appears once per layout as a stacked bar.}
  \label{fig_layout_comparison_results}
\end{figure}

\begin{figure}[t]
  \centering
  \includegraphics[width=0.93\linewidth]{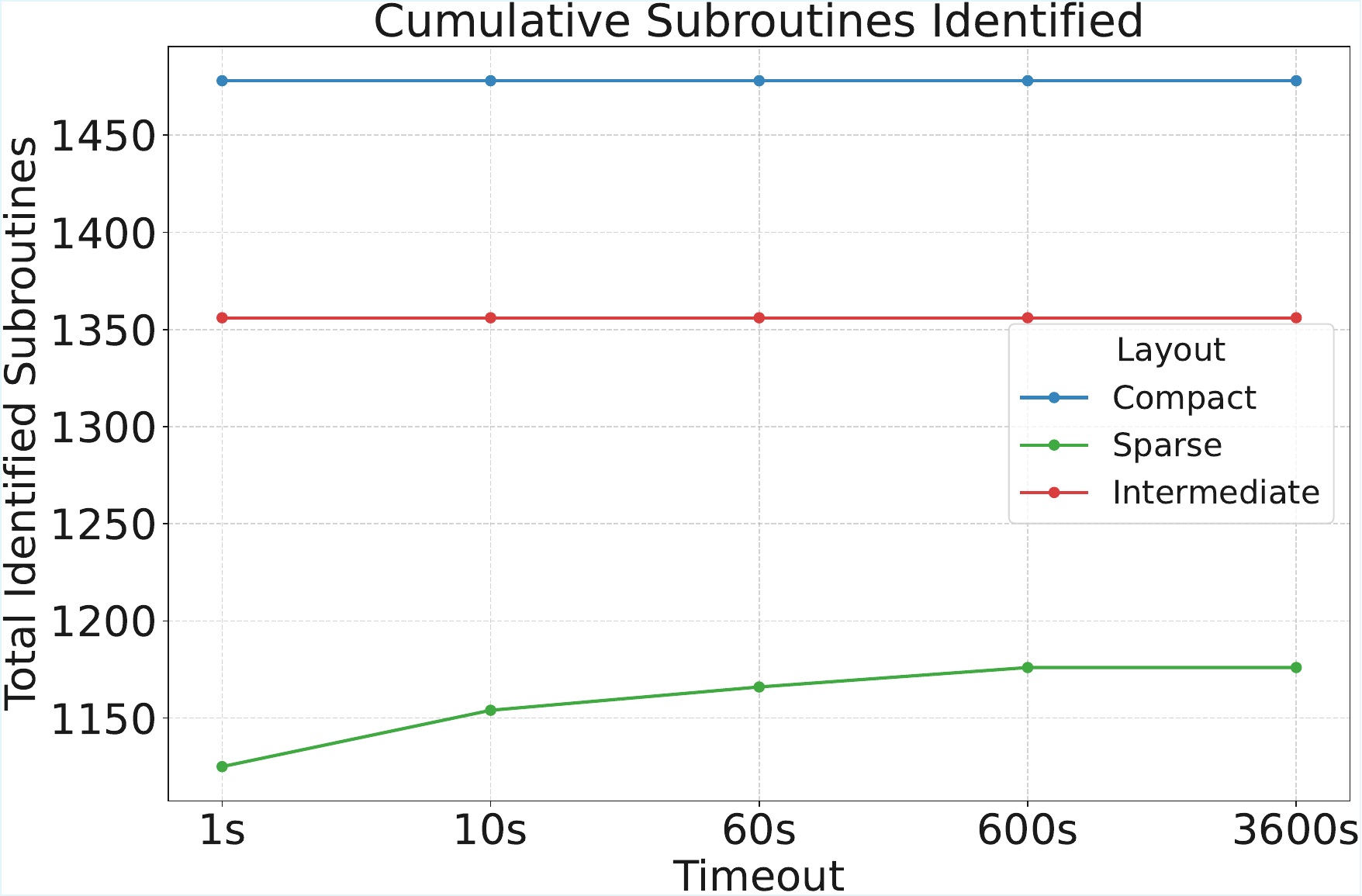}
  \caption{\small Cumulative number of identified subroutines across increasing timeout intervals for each layout. Each line shows how many successful subroutine matches were found within a given timeout limit for the compact, sparse, and intermediate layouts. Results are aggregated across all synthetic benchmarks and perturbations. The x-axis uses uniform spacing for clarity.}
  \label{fig_cumulative_subroutines_identified}
\end{figure}

Figure~\ref{fig_box_plot_ambiguity} shows how many trace entries per each synthetic benchmark, per each layout, are ambiguous. A trace entry is defined as ambiguous if there is at least one path which its directionality is unknown and can be more than one valid paths to connect the qubit endpoints. In all benchmarks, the Square Sparse layout produced strictly more ambiguous traces than the other layouts. The only exception was for synthetic benchmarks random\_mix\_9 and random\_mix\_10, where the synthetic benchmarks were naive in terms of qubit size and number of DAG nodes. This is mainly due to the fact that Square Sparse represents a case with a large space footprint. It also surrounds each logical qubit with one routing ancilla on each side. Consequently, it is highly likely to observe higher degree of parallelism for each benchmark, and thus more likely to observe higher ambiguity for valid paths between qubit endpoints.

For ambiguous trace entries, it is crucial to apply a traversal algorithm, such as Depth-First Search algorithm (DFS) with backtracking logic to identify all the possible valid paths among the recovered qubit endpoints. Figure~\ref{fig_boxplot_dfs_runtime} presents the runtime of the DFS algorithm for every synthetic benchmark per each layout. For Compact and Intermediate layouts, the DFS execution concluded in less than one second for every perturbation of each synthetic benchmark. For the Square Sparse layout, we observe higher execution times on average, while there are some outliers which need more than 100 seconds to conclude. This behavior is expected for Square Sparse, since this layout enables higher parallelism degree and also its grid dimensions are significantly larger than the two other candidate layouts. As a result, more trace entries are marked as ambiguous and path lengths are significantly larger, resulting in slower execution.

Figure~\ref{fig_post_dfs_success_rate_v5} depicts the success rate of recovering the qubit endpoints and their valid connection paths after applying the DFS algorithm. All DFS runs were allotted a 1 hour timeout, single-threaded, on an Intel(R) Xeon(R) Gold 6438N Processor and 256GB of RAM accessed via a distributed research cluster. After applying our heuristics to recover the L3 traces from original L1 traces, if during the DFS we found odd number of logical qubits recovered for a trace entry, we raise an exception and halt the execution. Every run concluded within the $1$ hour timeframe. We observe that Square Sparse layout appears to have the worst success rate on average. The lower the success rate, the fewer perturbations per synthetic benchmark were identified without any odd number of qubit endpoints. As expected, for the Square Sparse layout we observe the lower success rate on average, due to higher ambiguity percentage on trace entries. Another interpretation of the endpoint pairing success rate is that it may correspond to successfully inferring the layout topology. The only case where this rate does not reflect the success rate of fully recovering the layout, is when we observe even number of qubit endpoints (and thus the execution proceeds), although in reality there were more qubit endpoints remained unidentified. However, as shown in the following paragraphs, these cases are~limited.

As shown in Figure~\ref{fig_dfs_layouts}, there is a clear positive correlation between the total number of qubit endpoints with ambiguous routing paths per perturbation in a given layout and the time required for the DFS search to complete. Layouts with greater spatial routing freedom, such as the Square Sparse, tend to generate more ambiguous endpoint pairings and thus incur longer DFS resolution times. In practice, the Square Sparse layout, which offers the highest degrees of freedom, produced some instances with on the order of $600$ ambiguous endpoints, driving DFS runtimes to nearly up to $250$ seconds for a few cases, whereas more space restrictive, but faster, layouts such as Compact and Intermediate, typically saw under $100$ ambiguous endpoints and correspondingly sub-second DFS execution runtime. This relationship indicates that a layout’s routing complexity directly translates to greater computational overhead in trace analysis, echoing the broader evaluation observation that highly ambiguous traces yield larger number of valid paths, which in turn yields larger reconstructed circuit graphs and thus slower subsequent matching processes.

After identifying the qubit endpoints and all the valid, non-overlapping connections paths, the next step is to reconstruct the DAG of this synthetic benchmark. We anticipate that the reconstructed DAG of the synthetic benchmark will be augmented in terms of nodes and edges due to the ambiguity of trace entries. In general, the more ambiguous trace entries a benchmark has, the more augmented is the reconstructed DAG (and thus the slower the subgraph matching is). Figures~\ref{fig_layout_sparse_results}, ~\ref{fig_layout_compact_results} and ~\ref{fig_layout_intermediate_results} show how many subroutine instances were identified per each layout. All subgraph matching runs were allocated 1 hour timeout single-threaded.

Overall, we did not observe any false-positive subroutine matches. All reported detections corresponded to real inserted subroutines, since our exact matching prevents spurious matches. The main issue was false-negatives (missed detections) in a few highly ambiguous cases. It is clear from Figure~\ref{fig_layout_sparse_results} that Square Sparse layout yields the most cases, which they have been either failed to match, or the subgraph matching did not finish within the allocated time. Again, this is an expected behavior since this layout is the one that we see clearly higher ambiguity in the trace entries, due to higher parallelism and longer paths. This results either to larger DAGs, which are more computationally expensive to match, or to a few failed cases. The reason behind the failure is that while we are able to identify always even number of qubit endpoints (and did not halt the execution), in reality we missed some of them, usually two (or more) qubit endpoints remained unidentified. For Figures~\ref{fig_layout_compact_results} and ~\ref{fig_layout_intermediate_results} we do not observe any case to have been timed out. This is reasonable, as the reconstructed DAG of synthetic benchmark contains fewer node and edges to match due to lower ambiguity in the Compact and Intermediate layouts. For Figure~\ref{fig_layout_compact_results} we observe only six cases to fail due to missed identification of some qubit endpoints. For Figure~\ref{fig_layout_intermediate_results}, we observe only one failed case to match for the same reason. 

Figure~\ref{fig_cumulative_subroutines_identified} shows the cumulative number of identified subroutines per each layout. Different runtime intervals for the subgraph matching were tested. For Compact and Intermediate layouts, all subgraph matching processes concluded within a second. As expected, for Square Sparse, the majority of the cases concluded within one second, although more time is needed beyond the allocated timeframe for the subgraph matching to identify common subroutines, such as the Trotterization circuits with steps $5$ and $6$, t\_npe\_5 and t\_npe\_6, respectively. The reason is that these two subroutine circuits are the most computationally expensive since for the Square Sparse layout, the synthetic benchmarks that consist of these subroutines (i.e., random\_mix\_4, random\_mix\_7 and random\_mix\_8) exhibit the highest ambiguity percentage in trace entries (Fig.~\ref{fig_box_plot_ambiguity}), and also the highest average number of DAG nodes for the Square Sparse layout (Table~\ref{tab_synthetic-benchmark}).

\subsection{Evaluation Results Summary}

Our findings indicate that fast layouts (such as Compact and Intermediate) are correlated with a higher chance of discovering subroutines patterns. When the quantum computer architecture trades-off time over space, our framework still has reasonable chance to identify common subroutines, although the overhead may be higher and the failure rate slightly~increased. Overall, after the DFS analysis and the subgraph matching process, our aggregated results show that TraceQ successfully recovered 74\% of the subroutines (with no false positives introduced) for Square Sparse layout, 95\% for Compact layout and 88\% for Intermediate layout.

\section{Related Work}
\label{related_work}

To the best of our knowledge, this is the only work so far on defining and analyzing operation of FTQC computers and quantum programs based on space-time activity patterns due to lattice surgery. One recent and relevant work we found is by Erata et al.~\cite{erata2024quantum}, in which, in context of computer security, reconstructs NISQ quantum circuits from power traces of the operation of the quantum computer controller. They effectively capture which gates are executed based on the power traces and reconstruct the circuit. Their work could be a form of trace based analysis of NISQ operation, although they do not formalize the traces nor develop automated framework. Our work focuses on the FTQC setting, formalizes access traces, and provides a framework of efficient identification of the dataflow.

\section{Discussion and Future Directions}
\label{future_work}

Our work is the first to present a framework for generating traces based on capturing two-qubit gate operations and then analyzing them. This work builds a foundation for future trace-based analysis of fault-tolerant quantum computing. The trace-based approach can be naturally extended to include T gates. This can be done by tracing the connections as the paths are merged (and later split) when accessing magic T states. This could provide more information in the reconstructed graph, however, there may be higher computational costs and uncertainty in the reconstruction when routing to and from magic state qubits. We found that considering only two-qubit gates gives us important insights, without added overhead. Our framework could also be extended to work with lower-level lattice surgery operations. Since CNOT gates may be compiled as two lattice surgery measurements \cite{fowler2018low}, the framework could handle lower-level traces where there are two (or more) entries in the trace for each two-qubit operation. Our current work follows the prior standard in the literature \cite{10.1145/3720416}, but a lower-level extension is possible.

In terms of the TraceQ software, intelligent pruning of the DFS search space could be an extension to further improve performance. For the subgraph matching part, future work could investigate advanced variants of the VF3 algorithm, such as its parallel implementation~\cite{carletti2017introducing}. Recent advancements in exact subgraph matching, including GPU-accelerated versions of VF3, have demonstrated the ability to recover subgraphs of varying size from target graphs with over $300,000$ nodes, highlighting the potential of exact methods at scale~\cite{vf_gpu}.

\section{Conclusion}
\label{conclusion}

In this work, we introduced TraceQ, a novel framework for reconstructing the dataflow of quantum programs from access traces in fault-tolerant systems. By leveraging space-time patch activation patterns, TraceQ identifies logical qubit endpoints, reconstructs routing paths via ambiguity-aware heuristics, and recovers the underlying dependency DAG of the original circuit. Our evaluation on synthetic benchmarks highlights the effectiveness of this approach across various layout strategies, revealing key trade-offs between layout, trace ambiguity, reconstruction success, and subroutine detection.  The results demonstrate that meaningful program-level structure can be recovered from simple access traces. Moving forward, we envision that this framework enables performance and security study of fault-tolerant systems. All the code will be made public, so various researchers can use it.

\section{Acknowledgements}

This work is funded in part by NSF grant 2332406.
This work is also funded in part by the STAQ project under award NSF Phy-232580; in part by the US Department of Energy Office of Advanced Scientific Computing Research, Accelerated 
Research for Quantum Computing Program; and in part by the NSF Quantum Leap Challenge Institute for Hybrid Quantum Architectures and Networks (NSF Award 2016136), in part based upon work supported by the U.S. Department of Energy, Office of Science, National Quantum 
Information Science Research Centers, and in part by the Army Research Office under Grant Number W911NF-23-1-0077. The views and conclusions contained in this document are those of the authors and should not be interpreted as representing the official policies, either expressed or implied, of the U.S. Government. The U.S. Government is authorized to reproduce and distribute reprints for Government purposes notwithstanding any copyright notation herein.

\bibliographystyle{IEEEtranS}
\bibliography{refs, ctk}

\end{document}